\documentclass[sigconf]{acmart}
\usepackage{amsmath,amsfonts,bm}
\usepackage{booktabs} %
\usepackage{url}
\usepackage{graphicx}
\usepackage{array}
\usepackage{color}
\usepackage{makecell}
\usepackage{multirow}
\usepackage{xspace}
\usepackage{colortbl}
\usepackage{subcaption}

\usepackage{pifont}
\usepackage{graphics}

\usepackage[export]{adjustbox}

\newcommand{\ours}{Anchor-DR\xspace}

\newcommand{\edit}[1]{#1}

\newcommand{\start}[1]{\vspace{.3mm}\noindent{{\bf #1}.}}

\definecolor{gred}{RGB}{255,102,102}
\definecolor{gblue}{RGB}{51,102,255}
\definecolor{gyellow}{RGB}{244,180,0}
\definecolor{ggreen}{RGB}{15,157,88}
\definecolor{ggrey}{RGB}{115,115,115}
\definecolor{na}{gray}{0.9}
\definecolor{LightYellow}{RGB}{255,255,191}
\definecolor{OrangeRed}{rgb}{1.0, 0.27, 0.0}
\definecolor{midnightgreen}{rgb}{0.0, 0.29, 0.33}
\definecolor{darkgreen}{rgb}{0.0, 0.42, 0.24}

\AtBeginDocument{%
  \providecommand\BibTeX{{%
    \normalfont B\kern-0.5em{\scshape i\kern-0.25em b}\kern-0.8em\TeX}}}

\copyrightyear{2023}
\acmYear{2023}
\setcopyright{acmlicensed}
\acmConference[SIGIR '23] {Proceedings of the 46th International ACM SIGIR Conference on Research and Development in Information Retrieval}{July 23--27, 2023}{Taipei, Taiwan, China.}
\acmBooktitle{Proceedings of the 46th International ACM SIGIR Conference on Research and Development in Information Retrieval (SIGIR '23), July 23--27, 2023, Taipei, Taiwan}
\acmPrice{15.00}
\acmISBN{978-1-4503-9408-6/23/07}
\acmDOI{10.1145/3539618.3592080}

\definecolor{midnightgreen}{rgb}{0.0, 0.29, 0.33}
\definecolor{darkpink}{rgb}{0.91, 0.33, 0.5}
\definecolor{darkpink}{rgb}{0.91, 0.33, 0.5}

\begin{document}

\title{Unsupervised Dense Retrieval Training with Web Anchors}

\author{Yiqing Xie}
\affiliation{\institution{Carnegie Mellon University}\country{USA}}
\email{yiqingxi@andrew.cmu.edu}

\author{Xiao Liu}
\affiliation{\institution{Microsoft}\country{USA}}
\email{xiliu2@microsoft.com}

\author{Chenyan Xiong}
\affiliation{\institution{Microsoft}\country{USA}}
\email{chenyan.xiong@microsoft.com}

\begin{abstract}
In this work, we present an unsupervised retrieval method with contrastive learning on web anchors. 
The anchor text describes the content that is referenced from the linked page. This shows similarities to search queries that aim to retrieve pertinent information from relevant documents.
Based on their commonalities, we train an unsupervised dense retriever, \ours, with a contrastive learning task that matches the anchor text and the linked document.
To filter out uninformative anchors (such as ``homepage'' or other functional anchors), we present a novel filtering technique to only select anchors that contain similar types of information as search queries. Experiments show that \ours outperforms state-of-the-art methods on unsupervised dense retrieval by a large margin (e.g., by 5.3\% NDCG@10 on MSMARCO). 
The gain of our method is especially significant for search and question answering tasks. Our analysis further reveals that the pattern of anchor-document pairs is similar to that of search query-document pairs.\footnote{Code available at https://github.com/Veronicium/AnchorDR.}

\end{abstract}




\maketitle

\section{Introduction}
Dense retrieval matches queries and documents in the embedding space \cite{DPR,ORQA,ANCE}\edit{, which can capture the semantic meaning of the text and handle more complex queries compared to traditional sparse retrieval methods \cite{BM25}.
Due to the scarcity of labeled data in certain domains, including legal and medical, numerous recent studies have focused on unsupervised dense retrieval, which trains dense retrievers without annotations \cite{ICT,contriever,cocondenser,QExt}.}

\edit{One of the most common approaches of unsupervised dense retrieval is to design a contrastive learning task that approximates retrieval \cite{ICT,cocondenser,COCOLM,contriever,spider,SPAR,QExt}, yet it is nontrivial to construct contrastive pairs.
Most existing methods construct contrastive pairs from the same context,} such as a sentence and its context \cite{ICT}, or two individual text spans in a document \cite{COCOLM,cocondenser,contriever}. The relation between these co-document pairs is different from query-document pairs in search or question answering, where the query aims to seek information from the document.
LinkBERT \cite{linkbert} leverage text spans sampled from a pair of linked Wikipedia pages. However, such text spans are not guaranteed to have high relevance.
\edit{Few other methods train a model to generate queries from documents \cite{InPars,QExt}, but they either require large language models or huge amounts of training data.}

In this work, we present \ours, an unsupervised dense retriever \edit{that is trained on predicting the linked document of an anchor given its anchor text.}
The text on the anchor of hyperlinks typically contains descriptive information that the source document cites from the linked document, 
\edit{suggesting that the anchor-document pairs exhibit resemblances to query-document pairs in search, where the search query describes the information that the user is required from the relevant document.}
As a result, we present to train \ours to match the anchor text and its linked document with a contrastive objective.

\edit{Although the relation between anchor-document pairs is typically similar to that of search queries and relevant documents}, there also exist a large number of uninformative anchors.
For example, a web document may use anchor links to redirect to the linked document (e.g., ``homepage'' or ``website''). 
\edit{Such anchor-document pairs do not resemble the relation between search queries and documents} and may introduce noise to our model.
We thus design a few heuristic rules to filter out functional anchors, such as headers/footers or anchors in the same domain.
In addition, we train a classifier with a small number of high-quality search queries to further identify anchors containing similar types of information as real search queries.

Experiment results show that \ours outperforms state-of-the-art unsupervised dense retrievers by a large margin on two widely adopted retrieval datasets, MSMARCO \cite{MSMARCO} and BEIR \cite{beir} (e.g., by 5.3\% NDCG@10 on MSMARCO).
\edit{The improvement of \ours is most significant on search and question answering tasks, suggesting that compared to the contextual relation between co-document text spans \cite{cocondenser,contriever}, the referral relation between anchor-document pairs is more similar to the information-seeking relation between search query-document pairs.
We further present examples to show that anchor-document pairs indeed have similar patterns as query-document pairs.}

\section{Related Work}
\start{Dense Retrieval}
Dense retrieval is the technique of using dense vector representations of text to retrieve relevant documents \cite{Deerwester1990IndexingBL,DSSM}. 
With the development of pretrained language models \cite{DPR,BERT}, recent works have developed various techniques for dense retrieval, including retrieval-oriented pretraining \cite{COCOLM,cocondenser,contriever} and negative selection \cite{ANCE}.
While dense retrieval has exhibited remarkable effectiveness in contrast to traditional sparse retrieval approaches \cite{BM25}, its benefits are generally confined to supervised settings that involve an adequate amount of human annotations \cite{beir}.

\start{Unsupervised dense retrieval}
Previous work on unsupervised dense retrieval mainly adopts contrastive learning to model training.
ICT \cite{ICT} matches the surrounding context of a random sentence.
SPAR \cite{SPAR} uses random sentences as queries with positive and negative passages ranked by the BM25 score.
Co-condenser \cite{cocondenser}, COCO-LM \cite{COCOLM}, and contriever \cite{contriever} regard independent text spans in one document as positive pairs.
QExt \cite{QExt} further improves their work by selecting the text span with the highest relevance computed by an existing pretrained model.
A few other research works use neural models to generate queries, such as question-like queries \cite{InPars} or the topic, title, and summary of the document \cite{QExt}.
However, both works require a large-scale generation system.

\start{Leveraging web anchors in retrieval}
\edit{Web anchors have been widely applied to classic approaches for information retrieval \cite{AnchorAnalysis,Anchorcollaborative,AnchorContent,AnchorMining,AnchorStructure}. 
Recently, HARP \cite{hyperlinkPretrain} designs several pretraining objectives leveraging anchor texts, including representative query prediction or query disambiguation modeling.
ReInfoSelect \cite{ReinfoSelect} learns to select anchor-document pairs that best weakly supervise the neural ranker.
However, these methods either focus on classic bag-of-word modeling or apply a cross-encoder architecture that does not fit the setting of dense retrieval.}

\section{Methodology}
\edit{We present an unsupervised dense retrieval method that trains the model to match the representations of anchor text and its linked document.}
This section describes the contrastive learning task of anchor-document prediction and the anchor filtering process.

\subsection{Contrastive Learning with Anchor-Document Pairs}
\label{sec:ad_pred}
\edit{Based on the commonalities between anchor-document pairs and query-document pairs \cite{AnchorAnalysis,Anchorcollaborative,AnchorContent,AnchorMining,AnchorStructure}, we compute the representation of each anchor and document with our model, \ours, and trains it with} a contrastive objective of matching anchor text and its linked document:
\begin{align}
\mathcal{L}(a, d_{+})&=-\frac{\exp (sim(a, d_{+}) )}{\exp (sim(a, d_{+}) )+\sum_{d_{-}\in Neg (a)} \exp (sim(a, d_{-}) )} \\
sim(a, d)&=\langle f_\theta(a), f_\theta(d)\rangle,
\end{align}
where $f_\theta$ is our presented model, \ours, with T5 \cite{T5} as its backbone, the sequence embedding $f_\theta()$ is the embedding of the first token output by the decoder of \ours, $(a, d_{+})$ is the anchor text and its linked document, and $Neg (a)$ is the set of negative documents sampled from the whole dataset. 
In practice, we use BM25 negatives in the first iteration \cite{DPR} and use the negatives mined by \ours in the following iterations \cite{ANCE}.

In inference, we feed the query and all the documents into \ours separately and use the embedding of the first token in the decoder output as the sequence embedding.
Then we rank all the documents by their similarity to the query: $sim(q, d)=\langle f_\theta(q), f_\theta(d)\rangle$, where $f_{\theta}$ denotes \ours.

\subsection{Anchor Filtering}
\label{sec:filtering}
While some anchor-document pairs exhibit strong similarities with query-document pairs in search, others do not.
\edit{For instance, ``homepage'' or ``website'' and their linked documents hold entirely distinct relations with query-document pairs.}
Including these pairs in the training data may introduce noise to our model.
As a result, we \edit{first apply a few heuristic rules and then train a lightweight classifier} to filter out uninformative anchor text.

\start{Anchor filtering with heuristic rules}
We observe that a large number of uninformative anchors are functional anchors and these anchors mainly exist between pages within the same website. 
Consequently, \edit{we filter out anchor text that falls in the following categories: 
(1) \textit{In-domain anchors}, where the source and target page share the same domain; 
(2) \textit{Headers or footers}, which are detected by specific HTML tags, such as <header> and <footer>;} and 
(3) \textit{Keywords indicating functionalities}, which are manually selected from anchors with top 500 frequency. \footnote{We list the keywords in github.com/Veronicium/AnchorDR/blob/main/anchor\_filtering}

\start{Anchor filtering with query classifier}
We train a lightweight query classifier to learn the types of information that is typically contained in search queries \edit{about relevant documents}.
Specifically, we use the \textit{ad-hoc} queries provided by WebTrack \cite{WebTrack} as positive examples. \edit{These small number of queries are manually selected to reflect important characteristics of authentic Web search queries for each year.}
As for negative examples, we sample a subset of anchors before filtering by our rules, which has the same size as positive examples
We train the query classifier with the Cross-Entropy Loss:
\begin{align}
    \mathcal{L} = \sum_x \mathbf{1}_{Pos} \cdot log(g(x)) + \mathbf{1}_{Neg} \cdot log(1-g(x)),
\end{align}
where $g$ is a miniBERT-based \cite{minibert} model. After training the query classifier, we rank all the anchor text by the logits of the positive class (i.e., similarity to search queries) and only keep the top 25\%.

\begin{table}[t]
\centering
    \caption{The statistics of ClueWeb22 anchor training data.}
    \vspace{-0.1cm}
    \label{tab:data_stat}
\resizebox{\linewidth}{!}{ 
\begin{tabular}{c|c|c|c|c|c}
\hline
\multicolumn{3}{c|}{\# of docs}
& \multicolumn{3}{c}{\# of anchors}                       
\\ 
\cline{1-6}
Raw 
& \begin{tabular}[c]{@{}c@{}}After filt.\\ by rules\end{tabular}
& \begin{tabular}[c]{@{}c@{}}After filt.\\ by model\end{tabular} 
& Raw
& \begin{tabular}[c]{@{}c@{}}After filt.\\ by rules\end{tabular}
& \begin{tabular}[c]{@{}c@{}}After filt.\\ by model\end{tabular} 
\\ 
\hline
60.49M 
& 10.17M
& 3.97M  
& 117.11M
& 20.66M 
& 4.25M
\\ 
\hline
\end{tabular}
}

\end{table}

\begin{table}[t]
    \centering
        \caption{Unsupervised retrieval results on MSMARCO and BEIR under nDCG@10. The best result for each task is marked in bold. The best result among dense retrievers is underlined. We follow previous work \cite{contriever} and report the average performance on 14 BEIR tasks and MSMARCO (BEIR14+MM). The results of coCondenser and results with $\dagger$ are evaluated using their released checkpoints. \edit{The results of other baselines are copied from their original papers.}
        }
    \label{tab:beir}
    \resizebox{0.49\textwidth}{!}{
    \begin{tabular}{@{}l c | c c c c | c}
        \toprule
        Model ($\rightarrow$) & BM25 & coCondenser & Contriever & SPAR & QExt & \textbf{\ours} \\
  Training Data & - & MSMARCO & Wiki+CCNet & Wiki & Pile-CC & ClueWeb \\
  \# Training Pairs & - & 8.8M & 3M+707M & 22.6M & 52.4M & 4.25M \\
  \midrule
  \midrule
  MS MARCO      & 22.8 & 16.2 & 20.6 & 19.3 & 20.6 & \bf \underline{25.9} \\
  TREC-COVID    & 65.6 & 40.4 & 27.4 & 53.1 & 53.5 & \bf \underline{77.4} \\
  BioASQ        & \bf 46.5 & 22.7 & \underline{32.7$\dagger$} & - & - & 31.9 \\
  NFCorpus      & \bf 32.5 & 28.9 & \underline{31.7} & 26.4 & 30.3 & 30.8 \\
  NQ            & 32.9 & 17.8 & 25.4 & 26.2 & 27.2 & \bf \underline{33.6} \\
  HotpotQA      & \bf 60.3 & 34.0 & 48.1 & \underline{57.2} & 47.9 & 53.2 \\
  FiQA-2018     & 23.6 & \bf \underline{25.1} & 24.5 & 18.5 & 22.3 & 21.1 \\
  Signal-1M     & \bf 33.0 & 21.4 & \underline{25.0$\dagger$} & - & - & 20.9 \\
  TREC-NEWS     & 39.8 & 25.4 & 35.2$\dagger$ & - & - & \bf \underline{45.5} \\
  Robust04      & \bf 40.8 & 29.8 & 32.7$\dagger$ & - & - & \underline{40.1} \\
  ArguAna       & 31.5 & \bf \underline{44.4} & 37.9 & 42.0 & 39.1 & 29.1 \\
  Touch\`e-2020 & \bf 36.7 & 11.7 & 19.3 & \underline{26.1} & 21.6 & 25.0 \\
  CQADupStack   & 29.9 & \bf \underline{30.9} & 28.4 & 27.9 & 27.1 & 29.1 \\
  Quora         & 78.9 & 82.1 & \bf \underline{83.5} & 70.4 & 82.7 & 72.1 \\
  DBPedia-ent   & 31.3 & 21.5 & 29.2 & 28.1 & 29.0 & \bf \underline{34.1} \\
  SCIDOCS       & 15.8 & 13.6 & 14.9 & 13.4 & 14.7 & \bf \underline{15.9} \\
  FEVER         & \bf 75.3 & 61.5 & 68.2 & 56.9 & 59.7 & \underline{71.1} \\
  Climate-fever & \bf 21.3 & 16.9 & 15.5 & 16.4 & 17.7 & \underline{20.6} \\
  SciFact       & \bf 66.5 & 56.1 & \underline{64.9} & 62.6 & 64.4 & 59.4 \\ 
  \midrule
  BEIR14+MM     & \bf 41.7 & 33.4 & 36.0 & 36.2 & 37.0 & \underline{39.9} \\
  All Avg.      & \bf 41.3 & 31.6 & 35.0 & - & - & \underline{38.8} \\
  Best on       & \bf 9 & 3 & 1 & 0 & 0 & \underline{6} \\
        \bottomrule
    \end{tabular}}

\vspace{-0.3cm}
\end{table}

\section{Experiments}
In this section, we describe the experiment setups, compare \ours with baselines and ablations, and analyze its effectiveness.

\subsection{Experimental Setup}
\label{sec:setup}
We evaluate \ours on two public datasets: MSMARCO \cite{MSMARCO} and BEIR \cite{beir} for unsupervised retrieval, where we directly apply the methods to encode test queries and documents without supervision.
We report the nDCG@10 results following previous works \cite{contriever,QExt}.

\start{Training data} We train \ours on a subset of the ClueWeb22 dataset \cite{clueweb}. To preprocess the data, we first randomly sampled a subset of English documents with at least one in-link. 
After that, we use rules and then train a query classifier to filter out uninformative anchors, as introduced in Sec.~\ref{sec:filtering}.
\edit{Finally, we sample at most 5 in-links for each document.}
The statistics of the anchors and documents after each step of filtering are shown in Table ~\ref{tab:data_stat}. 
Note that ClueWeb22 has in total of 52.7B anchors, hence we are able to further scale up our model in the future.

\start{Implementation details} For continuous pretraining on anchor-document prediction, we train our model with BM25 negatives for one epoch and with ANCE negatives \cite{ANCE} for another epoch. We use a learning rate of 1e-5 and a batch size of 128 positive pairs. 
\edit{The query classifier is trained on the \textit{adhoc} test queries of WebTrack 2009 - 2014 \cite{WebTrack}, which contains 300 queries in total.}

\begin{table}[t]
    \centering
     \caption{nDCG@10 of models trained with different contrastive tasks on the same subset of documents, with 400K documents and 400K contrastive pairs. T-test shows \ours outperforms co-doc on All Avg. with p-value < 0.05.}
    \label{tab:ablation}
    \resizebox{0.4\textwidth}{!}{
    \begin{tabular}{@{}lccc|c}
        \toprule
        Model ($\rightarrow$)
        & ICT
        & co-doc
        & Anchor (rule only)
        & \textbf{\ours} \\
  \midrule
  \midrule
  MSMARCO      & 20.9 & 19.9 & 20.3 & \bf 22.1 \\
  TREC-COVID    & 65.0 & 64.4 & \bf 72.4 & 70.6 \\
  BioASQ        & 29.7 & 26.7 & 29.7 & \bf 30.8 \\
  NFCorpus      & 27.1 & 24.0 & 24.7 & \bf 28.4 \\
  NQ            & 23.4 & 27.9 & 30.7 & \bf 31.0 \\
  HotpotQA      & 39.8 & 38.9 & 41.5 & \bf 48.9 \\
  FiQA-2018     & \bf 20.4 & 17.8 & 19.1 & 18.2 \\
  Signal-1M     & 19.7 & 18.1 & 19.5 & \bf 21.1 \\
  TREC-NEWS     & 37.1 & 39.2 & \bf 43.3 & 42.5 \\
  Robust04      & 30.4 & 34.6 & 34.9 & \bf 38.2 \\
  ArguAna       & 39.7 & \bf 45.1 & 26.0 & 26.5 \\
  Touch\`e-2020 & 23.0 & 25.4 & \bf 27.4 & 25.2 \\
  CQADupStack   & 26.3 & 26.2 & \bf 26.7 & 24.8 \\
  Quora         & 76.6 & 74.9 & \bf 77.3 & 71.6 \\
  DBPedia-ent   & 25.2 & 26.7 & 27.5 & \bf 31.4 \\
  SCIDOCS       & 14.0 & 13.6 & 13.8 & \bf 14.7 \\
  FEVER         & 57.7 & 56.5 & \bf 72.2 & 69.8 \\
  Climate-fever & 19.5 & \bf 20.0 & 21.2 & 18.3 \\
  SciFact       & 54.1 & 54.3 & 50.4 & \bf 56.1 \\ 
  \midrule
  BEIR14 + MM   & 35.5 & 35.7 & 36.7 & \bf 37.2 \\
  All Avg.          & 34.2 & 34.4 & 35.7 & \bf 36.3 \\
        \bottomrule
    \end{tabular}
    }
\end{table}

\start{Baselines}
We compare \ours with a sparse retrieval method: BM25 \cite{BM25} and four unsupervised dense retrieval methods: coCondenser \cite{cocondenser}, Contriever \cite{contriever}, SPAR $\Lambda$ (trained on Wikipedia) \cite{SPAR}, and QExt-PLM (trained on Pile-CC with MoCo) \cite{QExt}. All these dense retrieval methods construct contrastive pairs in an unsupervised way: either by rules \cite{cocondenser,contriever}, lexical features \cite{SPAR}, or with pretrained models \cite{QExt}. 
Note that we do not compare with methods that require large-scale generation system to generate contrastive pairs, such as QGen \cite{QExt} or InPars \cite{InPars}, as their generators either require additional human annotations or have significantly larger sizes compared to our model (e.g., 6B vs. 220M).

As for ablation studies, we substitute the anchor-document prediction task with two other contrastive tasks: 
\textit{ICT} \cite{ICT}, which \edit{considers a document and a sentence randomly selected from the document as positive pairs}, and 
\textit{co-doc} \cite{cocondenser}, which \edit{treats two text sequences from the same document as positive pairs}.
We also compare to \textit{Anchor (rule only)}, which removes the query classifier and only uses rules to filter anchors.
For a fair comparison, we train all the ablations on the same subset of documents in ClueWeb22.

\begin{table*}[t]
\centering
\caption{Examples of the query-document pairs in two BEIR datasets: ArguAna and TREC-COVID, the co-document text pairs (co-doc), and the anchor-document pairs (\ours).}
\vspace{-0.25cm}
\label{tab:case}
\resizebox{2.12\columnwidth}{!}{
\begin{tabular}{p{20cm}}
    \toprule
    \underline{\textit{\textbf{Dataset}: ArguAna}}
    \hskip4em
    \textbf{Query}: Becoming a vegetarian is an environmentally friendly thing to do. Modern farming is one of the main sources of pollution in our rivers, and as long as people continue to buy fast food ...
    \hskip5.9em
     \textbf{Document}: Health general weight philosophy ethics	You don’t have to be vegetarian to be green. Many special environments have been created by livestock farming, for example chalk down land in England and mountain pastures ...
     \\ 
     
    \midrule
    \underline{\textit{\textbf{Dataset}: TREC-COVID}}
    \hskip2.2em
    \textbf{Query}: what causes death from Covid-19?
    \hskip2.7em
     \textbf{Document}: Predicting the ultimate outcome of the COVID-19 outbreak in Italy: During the COVID-19 outbreak, it is essential to monitor the effectiveness of measures taken by governments on the course of the epidemic. Here we show that there is already a sufficient amount of data collected in Italy to predict the outcome of the process ...
     \\ 

    \midrule
    \midrule
    \underline{\textit{\textbf{Method}: Codoc}}
    \hskip5.4em
    \textbf{Query \#1}: Going vegetarian is one of the best things you can do for your health.
    \hskip4em
     \textbf{Document \#1}: We publish a quarterly magazine The Irish Vegetarian, with features and our roundup of news and events of interest to Irish vegetarians. 
Get involved! There are lots of ways to get involved. 
You can read our Going Vegetarian page. You can pick up a copy of The Irish Vegetarian. You can come to a Meetup meeting ...
     \\ 
     \\
    \textbf{Query \#2}: COVID-19 vaccines designed to elicit neutralizing antibodies may sensitize vaccine recipients to severe diseases
    \hskip4em
     \textbf{Document \#2}: According to a study that examined how informed consent is given to COVID-19 vaccinetrial participants, disclosure forms fail to inform volunteers that the vaccine might make them susceptible to more severe disease. The study, “Informed Consent Disclosure to Vaccine Trial Subjects of Risk of COVID-19 Vaccine ...
     \\ 
     
    \midrule
    \underline{\textit{\textbf{Method}: \ours}} \hskip3.4em
    \textbf{Query \#1}: Vegetarian Society of Ireland
    \hskip4em
     \textbf{Document \#1}: The Vegetarian Society of Ireland is a registered charity. Our aim is to increase awareness of vegetarianism in relation to health, animal welfare and environmental perspectives. We support both vegetarian and vegan aims. Going vegetarian is one of the best things you can do for your health, for animals and for the planet ...
     \\ 
     \\
    \textbf{Query \#2}: How COVID19 Vaccine Can Destroy Your Immune System
    \hskip4em
     \textbf{Document \#2}: According to a study that examined how informed consent is given to COVID-19 vaccine trial participants, disclosure forms fail to inform volunteers that the vaccine might make them susceptible to more severe diseases...
     \\ 
    \bottomrule
\end{tabular}
}
\vspace{-0.25cm}
\end{table*}

\subsection{Main Results}
\label{sec:main}
Table \ref{tab:beir} shows the unsupervised retrieval results on MSMARCO and BEIR. \ours outperforms all the dense retrieval baselines on MSMARCO and BEIR with a large margin (e.g., by 2.9\% nDCG@10 on BEIR14+MM and 3.8\% on all datasets).
Furthermore, compared to other dense retrievers, \ours achieves the best performances across a majority of datasets.
indicating that our method can be generalized to a wide range of domains and retrieval tasks.

\edit{We observe that \ours exhibits strong performance in specific subsets of tasks. For instance, \ours achieves a large performance gain of 11.8\% nDCG@10 on TREC-COVID, but it is outperformed by other baseline methods on ArguAna and Quora.}

\subsection{Ablation Study}
\label{sec:ablations}
To demonstrate the effectiveness of our anchor-doc prediction task, we perform ablation studies in Table \ref{tab:ablation}.
We observe that \ours outperforms both methods.
Additionally, \textit{ICT} and \textit{co-doc} have less than 1\% performance gap on 7 out of 19 datasets. This is probably because the contrastive learning pairs in both methods contain contextual information about each other.
\ours also outperforms \textit{Anchor (rule only)}, indicating that it is effective to train on anchor texts with higher similarities to search queries.

\begin{figure}[h]
  \centering
  \subfloat[Task Type]{
    \includegraphics[height=4.8cm,valign=t]{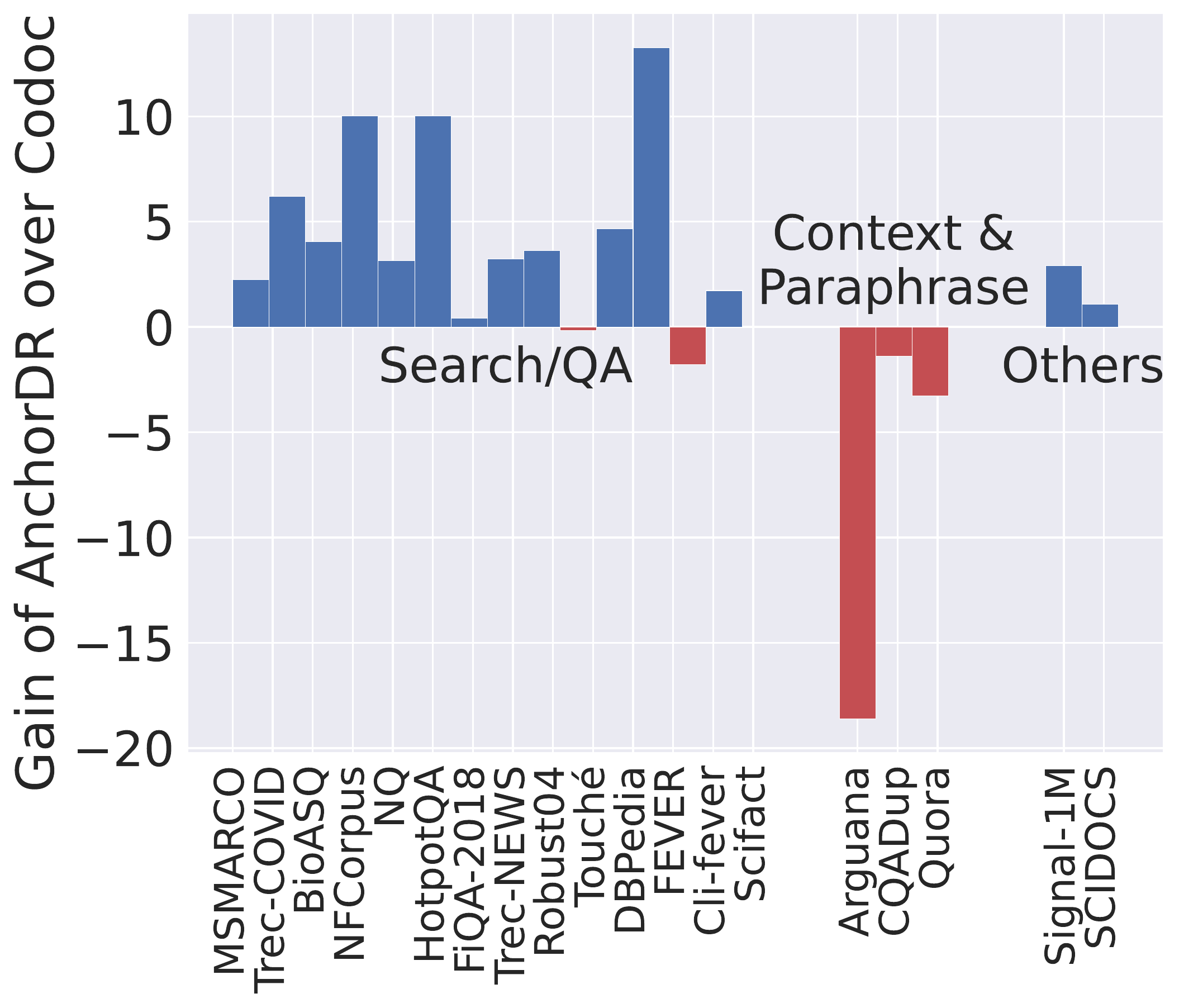}
  }
  \raisebox{0.0cm}{\subfloat[Lexical Overlap]{
    \raisebox{4.45cm}
    {\includegraphics[height=4.1cm,valign=t]{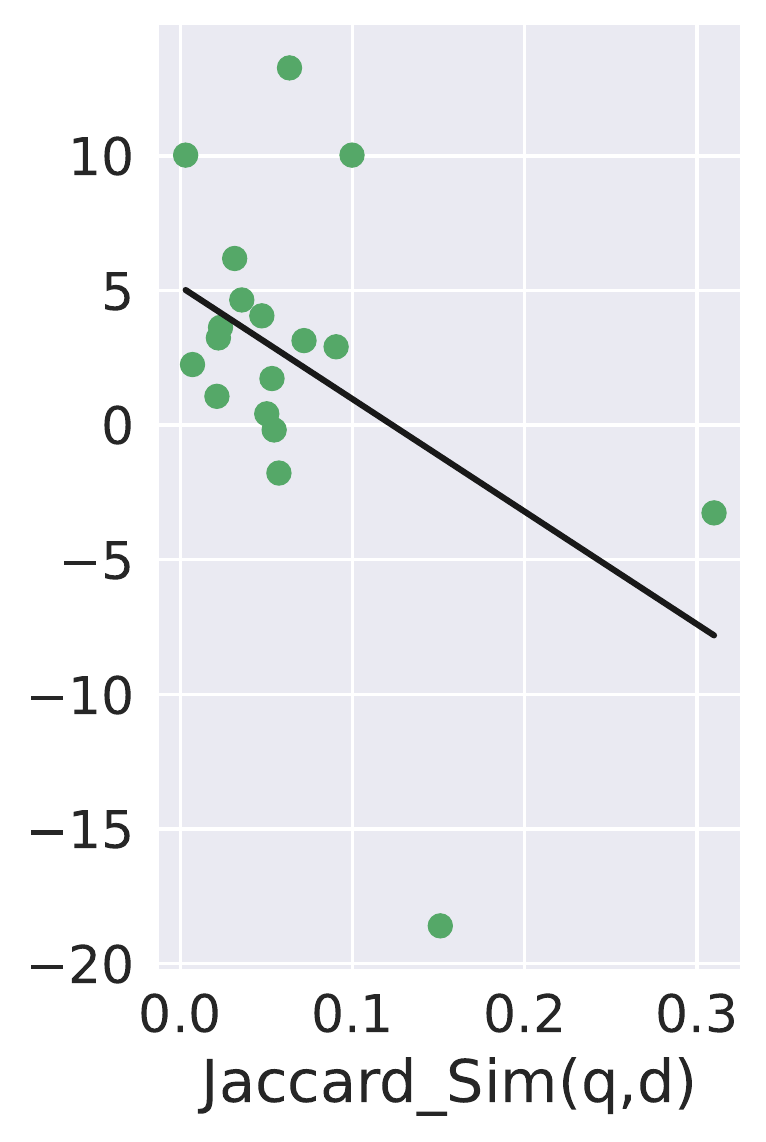}
  }}}
  \caption{Performance gain of \ours over codoc on different datasets under nDCG@10.}
\vspace{-0.5cm}
\label{fig:breakdown}
\end{figure}

\subsection{Performance Analysis}
\label{sec:analysis}
\start{Performance breakdown}
The results in Table \ref{tab:beir} show that \ours achieves strong performance in a majority of datasets but not in others.
To analyze the effectiveness of \ours on different datasets, we categorize the datasets into three subsets: (1) \underline{Search/QA}, where the query is a question or keywords related to the document; (2) \underline{Context/Paraphrase}, where the query and document contain coherent or overlapping information; and (3) \underline{Others}.
Figure \ref{fig:breakdown}(a) shows that \ours performs better on Search/QA datasets and co-doc is better on Context/Paraphrase datasets. 
\edit{The results are consistent with our hypothesis that the referral relation between query-document pairs is similar to the information-seeking relation between search queries and relevant documents.}

We further quantitatively analyze the information pattern of query-document pairs captured by \ours and co-doc.
Figure \ref{fig:breakdown}(b) shows the performance gap between \ours and co-doc versus the degree of information overlap between queries and documents in each test dataset, which is measured using Jaccard Similarity.
We observe that \ours performs much better on datasets where queries and documents contain less overlapping information. 
\edit{The primary emphasis of datasets with high query-document similarity is mainly on paraphrasing and coherency, which are distinct from the relation between search queries and documents.}

\start{Case studies}
We show in Table \ref{tab:case} the contrastive pairs of \ours and co-doc, as well as the positive pairs in ArguAna and TREC-COVID, which represent the Search/QA and Context/Paraphrase datasets.
The query-doc pairs of ArguAna are arguments around the same topic, which are coherent and have similar formats.
Similarly, the contrastive pairs of co-doc contain either coherent (e.g., the claim and recent work of the vegetarian society) or repeating information (e.g., COVID vaccine may cause diseases), which may explain its good performance on Context/Paraphrase datasets.

In contrast, in TREC-COVID, the answer to the query is contained in the document.
As shown in Table \ref{tab:case}, the anchor text in \ours could be the topic of the linked document, or in the format of a question. 
In both examples, \edit{the anchor text can serve as a search query and the document can provide the information the query is seeking},
which could be the reason why \ours achieves strong performance on the Search/QA datasets.

\section{Conclusion}
We train an unsupervised dense retrieval model, \ours, leveraging the rich web anchors.
In particular, we design a contrastive learning task: anchor-document prediction to continuously pretrain \ours.
Additionally, we apply predefined rules and train a query classifier to filter out uninformative anchors.
Experiments on two public datasets: MSMARCO and BEIR show that \ours significantly outperforms the state-of-the-art dense retrievers on unsupervised retrieval. Our analyses provide a further comparison of the patterns of information contained in our contrastive learning pairs and query-document pairs in test datasets.

\bibliographystyle{ACM-Reference-Format}
\bibliography{sample-base}

\appendix

\end{document}